\documentclass[preprint,aps]{revtex4}

\usepackage{graphicx}% Include figure files
\usepackage{dcolumn}% Align table columns on decimal point
\usepackage{bm}% bold math

\begin{document}

\title{ Quantitative Evaluation of the Effects of Positional versus Orientational 
Disorder on the Scattering of Acoustic Phonons in Disordered Matter}

\author{F.J. Bermejo, R. Fern\'andez-Perea and C. Cabrillo}%
\email{jbermejo@we.lc.ehu.es}
\affiliation{Instituto de Estructura de la Materia, C.S.I.C., and
Dept. Electricidad y Electr\'onica-Unidad Asociada CSIC,
Facultad de Ciencia y Tecnolog\'{\i}a, Universidad del Pa\'{\i}s Vasco / EHU,
 P.O. Box 644, E-48080-Bilbao, Spain}
\author{A.I. Krivchikov, A.N. Yushchenko, V.G. Manzhelii and O.A. Korolyuk}
 \affiliation{B.Verkin Institute for Low Temperature Physics and Engineering of NAN
Ukraine Kharkov, Ukraine}
\author{M.A. Gonz\'alez, M. Jimenez-Ruiz}
\affiliation{Institute Laue Langevin, 6 Rue Jules Horowitz,
F-38042-Grenoble Cedex 9, France}

\date{\today}

\begin{abstract}
The phonon scattering processes on the three solid phases of ethanol 
are investigated by means of thermal conductivity, light and neutron scattering 
measurements as well as molecular 
dynamics simulations on single-crystalline models for the two crystalline modifications 
(fully ordered monoclinic  and orientationally disordered $bcc$ phases). The orientationally 
disordered crystal is found to exhibit a temperature dependence of the thermal conductivity that
is  remarkably close to that found for the structurally amorphous solid, specially at low temperatures.
The results, together with measurements of the Brillouin linewidths as 
derived from light scattering measurements emphasize the role of orientational disorder on 
phonon scattering.  The experimental results obtained on  polycrystal 
samples are then discussed with the aid of computer simulations on  single-crystalline models of 
both $bcc$ and monoclinic crystals.  Our findings are in good agreement with the wealth of 
thermodynamic and dynamic data available so far \cite{eto1}, but are in stark contrast with 
inferences made from the analysis of inelastic X-ray data on polycrystalline 
samples \cite{matic}, where a common nature for the excitations in all phases is postulated. 

\end{abstract}

\maketitle

\section{Introduction}

The present understanding of the mechanisms governing heat transport in
disordered media rests upon concepts grounded on  
heat-pulse experiments showing that acoustic phonons, specially those having transverse
polarization are the main heat carriers \cite{vu}. The lifetime
and therefore the mean-free-path of such excitations is known to be
severy limited due to the action of several scattering mechanisms 
which are additional to those mainly dealing with the kinematics of the 
phonon gas, that are well established for fully ordered crystals.
 
One of the most striking findings revealed from work
carried out over the last couple of decades, concerns the
{\it quantitative} similarities exhibited by the thermal
conductivity of bulk amorphous materials \cite{rev} between say
0.1 K and 10 K, independent of chemical composition. In fact,
from the collected dataset, which now includes a good number of 
disordered crystals, including a quasicrystal \cite{rev,quasi},
 it is found that  the ratio of the
wavelength $\lambda$ of the acoustic wave to the mean free path
$\it l$  of all these solids ranges within 10$^{-2}$ - 10$^{-3}$,
which suggests the presence of `'universal'' behavior of some
sort. On such grounds, it is becoming increasingly clear that the presence of
`'glassy dynamics'' cannot be attributed in full to the absence of
static translational long-range order (LRO).

Within the disordered crystals referred to above, some molecular or
ionic materials where the constituent particles have random
static orientations while their centers of mass sit at the nodes
of a three dimensional crystalline lattice are also known to
exhibit glass-like excitations. Within those, solid ethyl alcohol
is perhaps the most convenient benchmark to carry out a
quantitative comparison of the effects brought forward by the
complete lack of LRO \cite{eto1} on the most sensitive property to
explore the propagation of excitations in condensed matter, such
is the thermal conductivity. The material, apart from the well
known $Pc$ monoclinic crystalline (fully ordered, Z = 2, FOC) modification,
can be prepared in three long-lived (although metastable) phases, 
that are an amorphous
solid or glass, an orientationally-disordered crystal, ODC (or
orientational glass) showing static orientational disorder but
having translational LRO since the molecules sit at the nodes of a
$bcc$ lattice, and a crystal with dynamic orientational disorder
(rotator-phase crystal or RPC) which retains LRO as a $bcc$
lattice still exists. Two calorimetric glass-transitions \cite{eto1}
take place about the same range of temperatures and centered about 97 K and correspond to
the canonical glass$\longrightarrow$ supercooled liquid and a 
rotational freezing transition
ODC$\longrightarrow$RPC \cite{eto1}, that is  well understood
as a purely dynamic phenomenon (see M.Jimenez-Ruiz {\it et al.} in 
Ref. \cite{eto1}).

Here we report on measurements of the thermal conductivity of
ethyl alcohol for all the solid phases as well as on measurements
of the acoustic phonon lifetimes as determined by means of optical 
Brillouin scattering.  Our aim in pursuing such a route
is twofold. First and foremost, as stated in a recent
review \cite{rev} the measurements will provide additional tests
on claims of quantitative `'universality'' on the properties of heat
propagation at low and intermediate temperatures on disordered
matter brought forward by a class of materials for which no
data were available in Ref.\cite{rev}. On a more fundamental vein,
our study also aims to verify recent claims concerning high-frequency 
(i.e. THz) acoustic excitations sampled in disordered matter by means of
inelastic X-ray (IXS) or neutron scattering (INS) \cite{silica,lb}.
More specifically, our results may contribute to clarify whether the
spectral feature characteristic of the INS or Raman spectra of most glasses
usually referred to as the `'Boson peak'' can be taken as a signature 
of a high-frequency limit beyond which acoustic excitations become 
localised as proposed by some authors \cite{silica,lb}, or contrary to
that, phonons may propagate down to microscopic scales without having
their lifetimes limited by strong scattering processes, as known to be 
the case for excitations up to GHz frequencies. In this respect, here 
we will particularly focus onto some details concerning   
the nature of the spectral response
observed by means of IXS on polycrystals of the same material
\cite{matic}, for which two rather different interpretations have been proposed
\cite{matic,lb}.  To the ends just delineated, we complement 
the experimental study with calculations of the phonon frequencies
and linewidths for both the FOC and ODC carried out by means of 
Molecular Dynamics simulations on single-crystalline models of both
materials using for the purpose well contrasted intermolecular 
potentials \cite{miguel}. In doing this we will follow the evolution
of the coherent dynamic structure factor $S(Q,E)$ accross the first 
Brilloin zone deriving unambiguous estimates for the phonon frequencies and
inverse lifetimes, an exercise which cannot be pursued in 
polycrystalline samples as known since long \cite{dewette}. 
Such calculations are complementary to previous harmonic Lattice Dynamics
results on the $Pc$ monoclinic phase carried out for a semi-rigid 
model which incorporates the lower lying molecular modes and
has been tested against thermodynamic (i.e. specific heat data up to
30 K) and spectroscopic data such as the frequency 
distributions as derived from neutron scattering as well as
Raman spectral data. For more details in this particular
aspect the interested reader is referred to Ref.\cite{ld}.

\section {Experiments and computational details}

Thermal conductivity measurements which are detailed in Ref.\cite{sasha}
were carried out under equilibrium vapor pressure at 2-159 K by the 
steady-state potentiometric method 
using a special setup \cite{Krivchikov-2005}. The 
preparation of the glass, disordered crystal or fully ordered 
monoclinic phase was carried out following previously described 
procedures \cite{eto1}

Brillouin light-scattering measurements were carried out using a 
multipass tandem Fabry-Perot interferometer from about 5 K up to 160 K
on backscattering geometry on both glass/liquid and ODC/RPC phases. 
Measurements on the monoclinic modification were hampered by the 
strong speckle patterns arising from the sample.

Neutron inelastic scattering experiments were performed using the
MARI spectrometer at ISIS (Rutherford Appleton Laboratory), using an
incident energy of 15 meV and a temperature of 10 K on fully 
deuterated samples. The different sample states were 
prepared in situ, the transformation between themmonitored by means of
diffraction measurements. Additional measurements on the Boson peak 
were carried out using teh IN14 triple axis spectrometer at the 
Institut Laue Langevin.

Molecular Dynamics simulations were carried out using a semi-flexible
representation of the molecules and its predictive capabilities are
described in Refs.\cite{miguel}.

\section{Results}

A summary of the measurements of the thermal conductivity of FOC, ODC
and RPC phases is shown in the Fig.1. A glance to such a figure shows

\begin{figure}
\includegraphics[width=3.1in,angle=0]{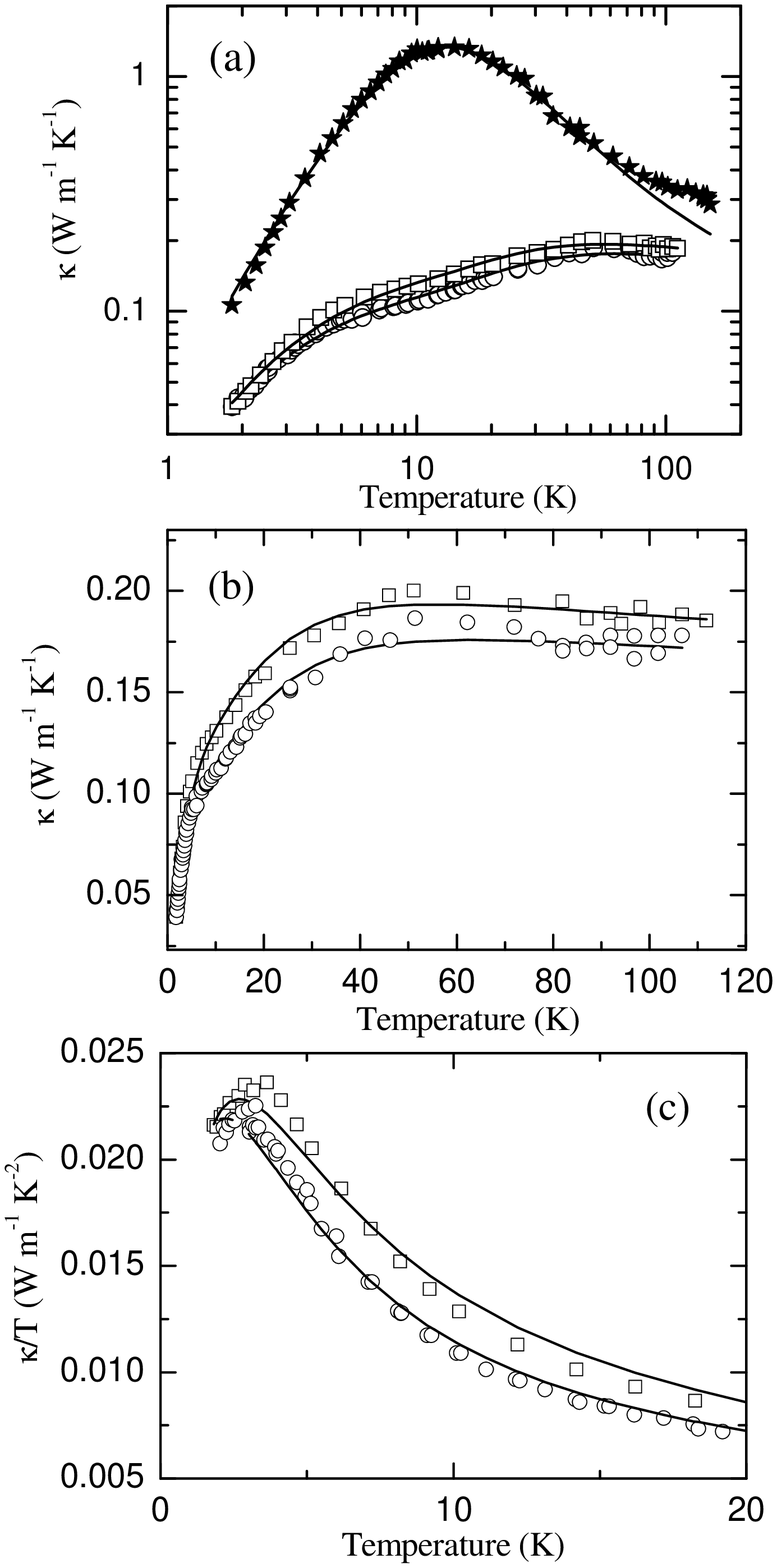} 
\caption{ The upper frame depicts the measured thermal conductivity 
for the FOC (stars), ODC (squares) and glass (circles). Solid lines 
are model fits as described in the text. The lower frame depicts data
for the glass (circles) and ODC (squares) plotted on a linear scale.}
\label{Fig.1} 
\end{figure}

that the conductivity of the FOC phase exhibits a shape much alike that of 
most crystalline solids explored so far. From melting ($\rm T_m=$ 159 
K) down to some 27 K, the thermal conductivity 
follows the exponential relation $\rm \kappa(T) \propto \exp{(-E/T)}$ with E = 38.3 
$\pm$ 0.9 K. The maximum is located at T = 12 K and below 4.5 K it then decreases 
following the well established  $\rm \kappa(T) \propto T^2$ behavior.
Notice the abrupt decrease in thermal conductivity by 42$\%$ in a narrow 
temperature interval. 
 
	  Both ODC and glass exhibit a temperature dependence of $\kappa$(T) 
rather similar to that of most amorphous solids \cite{rev}. The thermal 
conductivity increases with temperature with the largest growth 
registered within the region below 4 K. A smeared out `'plateau'' is 
then found within 5 K - 10 K, above which $\rm \kappa(T)$ experiences 
a further increase. Such an increase lasts up T = 50 K, from where  $\kappa$(T)
for both samples shows a very mild dependence on temperature up to both
glass-transitions range at $\rm T_g \simeq$ 97 K. The data for the ODC 
closely follows that of the amorphous solid. In real numbers, the thermal conductivity
of the ODC is higher than that of the structural glass 
by about 2 per cent at 2 K, 8 per cent at 3.2 K, 16 per cent at 10 K 12 
per cent at 25 K and 85 per cent above 50 K. At temperatures beyond $\rm 
T_g$ our data shows that $\rm \kappa(T)$ for the
supercooled liquid and for the RPC are both independent of temperature, its 
value being close to the thermal conductivity of liquid ethanol at the temperature 
of solidification, $\rm T_m$ = 159 K.

 The data has been analyzed on phenomenological grounds by means of 
 the model of soft potentials \cite{sasha}, where phonon scattering is 
 pictured as  caused by low Ð energy excitations such as tunneling states,
classical relaxors and soft quasiharmonic vibrations. As the most 
relevant parameters derived from fitting $\kappa(T)$ to such a model 
one gets estimates for the dimesnionless tunneling strength $C$ as well
as the quentity $W$ which characterizes the crossover from 
a regime dominated by phonon scattering by low-energy excitations (tunneling states and
classical relaxors) into another dominated by scattering by soft quasiharmonic vibrations.
Numerical values for both quantities yield 8.0$\times 10^{-4}$, 8.6$\times 
10^{-4}$ and 4.5$\times 10^{-4}$ for the 
glass, ODC and FOC respectively, while those for $W$ yield 3.6 K, 4.0 
K and 14.0 K for the same three phases. Put into different words, both 
the tunnel strenght and the crossover energy yield very close values
for both glass and ODC but largely differ from that found for the
fully ordered crystal.

From values found for the strength parameter $C$ a value for the ratio of 
the wavelength of the acoustic wave to its 
mean-free-path  the relationship given by Pohl {\it et al.} \cite{rev}
a value for the ratio of the wavelength of the acoustic wave to its 
mean-free-path is derived,
\begin{equation}
\frac{\lambda}{\it l}=\pi^2 C,
\end{equation}
and yields figures $\simeq$ 8$\times$ 10 $^{-3}$, that are in quantitative
agreement with data reported in Ref.\cite{rev} and thus lend further support to the claim
of `'universal'' behavior there expounded since it is shown to be also 
followed by this class of materials. Notice that the estimate for the 
monoclinic crystal would come close to an order of magnitude larger 
and thus well outside charateristic limits of disordered matter.

A direct experimental route to asess the findings just referred to is
provided by measurements of the phonon frequencies and linewidths by 
means of Brillouin light-scattering spectroscopy. Notice that for a 
disordered solid, it is within this limit where the concept of a
phonon retains its physical meaning. This comes as a consequence of
the fact that at scales where the experiments are typically performed that is
 wavevectors of the order
of 10$^{-3}$ \AA$^{-1}$ and frequencies of some GHz that are 
comparable to those sampled by thermal conductivity measurements, the 
solid behaves as an elastic continuum where acoustic excitations 
exhaust the frequency spectrum. 

The experimental spectra consist on a single peak attributable to 
longitudinal phonons, the transverse component being too weak to be
observed. The quantities of interest are here the Brillouin 
peak frequencies $\omega_{B}$ as well as the peak widths $\Gamma_{B}$.
From such quantities we calculate the fractional frequency shift $\Delta_{B}$ 
and the ratio of Brillouin linewidth to its frequency $Q^{-1}$ as
\begin{equation}\label{brill_parms}
\Delta_{B}=\frac{\delta \omega_{B}}{\omega}\; 
Q^{-1}=\frac{\Gamma_{B}}{\omega_{B}}.
\end{equation}
Here, the fractional frequency shift is calculated with respect to 
the datum measured for T = 5 K and provides a measure of the shift 
with temperature of the phonon spectrum, that is an estimate of the 
temperature coefficient of anharmonicity. In turn, $Q^{-1}$ provides 
an estimate of phonon damping in terms of a coefficient of internal 
friction.
 
\begin{figure}
\includegraphics[width=3.1in,angle=0]{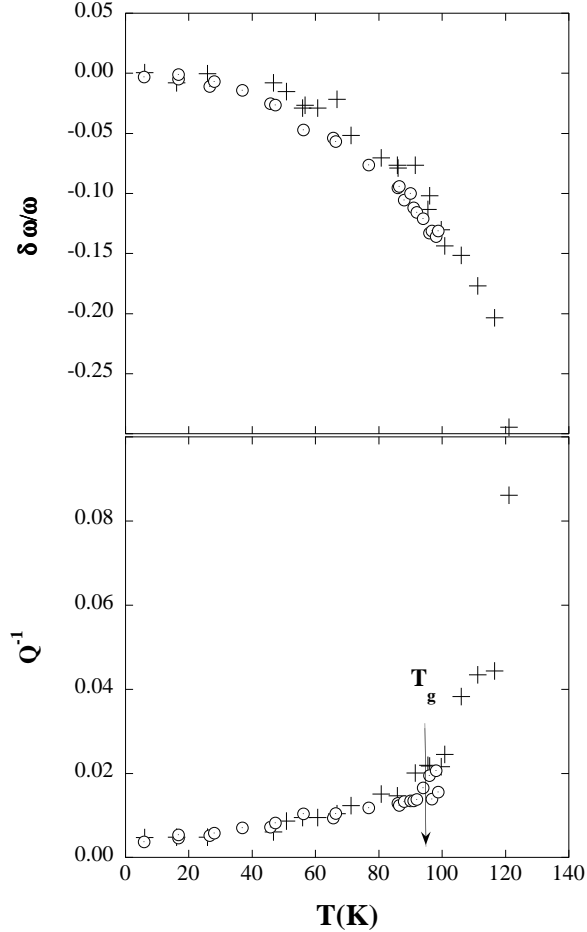} 
\caption{ The upper frame depicts the measured temperature dependence 
of the fractional change in sound velocity for  ODC/RPC (circles with 
a dot) and glass/liquid (crosses).}
\label{Fig2} 
\end{figure}

The Fig.\ref{Fig2} displays a comparison of the evolution with 
temperature of both $\Delta_{B}$ and $Q^{-1}$ for both glass and ODC
and temperatures covering most of the range of existence of both solids.
The behaviour with temperature of $\Delta_{B}$ for both solids follow
a rather mild dependence below some 30 K - 40 K which gets largely
enhanced at higher temperatures. In contrast, data for the internal 
friction follows nearly the same quasilinear behaviour for both
solids up to temperatures close to $T_{g}$. Both measures of the
anharmonic interactions in the two solids show remarkably close 
values, which put into different words means that the phonon
scattering processes operative in both kinds of disordered systems
are basically the same, which is in agreement with the thermal 
conductivity results described above. The results contrast with the
expectancy of a far less anharmonic behavior at low frequencies 
for the FOC. In fact macroscopic estimates of the $\gamma_{-3}$
Gr\''uneisen coefficient (see H.E. Fischer {\it et al.} in Ref.\cite{eto1})
for low frequencies yield values of 2.9, 4.0 and 4.4 for FOC, ODC and 
glass respectively.

\begin{figure}
\includegraphics[width=2.5in,height=3.6in ,angle=0]{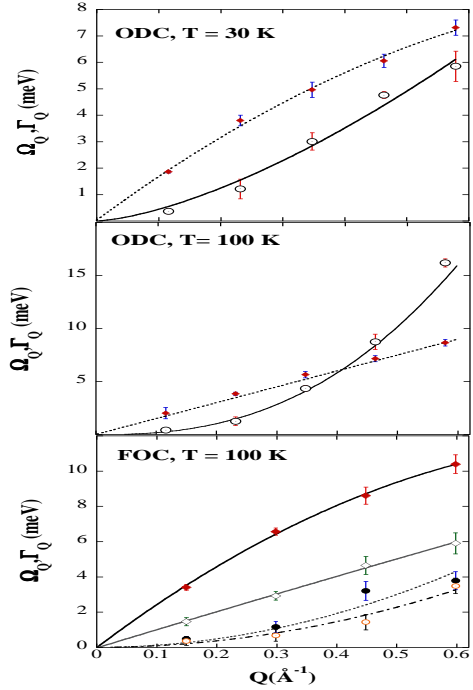}
\caption{Calculated phonon frequencies, $\Omega_{Q}$ and phonon linewidths
$\Gamma_{Q}$ along the (001) crystal directions. The upper frame compares
data for the orientationally disordered crystal (ODC) at low (5 K) and moderate (30 K)
temperatures. Thin and open crosses display  $\Omega_{Q}$ and
$\Gamma_{Q}$ the longitudinal phonon at T = 5 K. Full and open 
spares-with-a-cross depict frequencies and damping for the transverse
excitation sampled at the same temperature. Filled lozenges and open 
circles-with-a-dot show dispersion frequencies and damping terms
for the longitudinal phonon at 30 K. The lower frame compares data
for the FOC and ODC crystals at high temperature (100 K). Longitudinal
and transverse phonon frequenecies for the FOC are shown by filled and
open lozenges respectively. The corresponding linewidths are depicted
by filled and open circles. Data for the frequencies and widths of the 
ODC crystal at the same temperature shown by crosses and open
crosses respectively. Lines drawn through the points are approximations
given in terms of simple power laws (see text).} \label{FigMD}
\end{figure}

To deepen into the origin of the differences in behavior between the
two crystals, we have carried out a number of Molecular Dynamics
simulations on models of both ODC and FOC single-crystals using
for the purpose potentials and algorithms already tested against
experimental data \cite{miguel}. The calculated dynamic structure
factors along the main (001, 010 and 100) crystal directions and temperatures
comprising  5 K - 30 K were analyzed in detail. A sample of the results is shown in
Fig.\ref{FigMD}.   The lowest temperature ( 5 K) corresponds to the upper limit of the strong
rise with temperature of $\rm \kappa(T)$, T = 30 K is the limiting
temperature where the Debye model is able to account for the
measured specific heat \cite{ld}, or put into different words,
where the specific heat can be fully accounted for in terms of
acoustic excitations only. Finally, T = 100 K comes close to the stability 
 limit of the ODC and also to the actual temperature where IXS measurements
 were conducted.

The simulation data for the ODC show that for T > 5 K
only the longitudinal acoustic phonon survives the strong scattering processes induced by
orientational disorder. The graphs shown in the upper frame of
Fig.\ref{FigMD} for the lowest temperature depict a longitudinal branch that corresponds to a
hydrodynamic sound velocity of 2639 m s$^{-1}$ that remains as a
well defined peak all along the Brillouin zone. Damping of such
excitation follows a subquadratic dependence with wavevector, the
linewidth being well approximated by $\Gamma_{Q} = 9.8 \;
Q^{3/2}$ \AA$^{-3/2}$. A transverse excitation also appears
within the first two Brillouin zones. Its linear dispersion
corresponds to a propagation velocity of 989 m s$^{-1}$ while
its large damping is accounted by a stronger than quadratic 
dependence $\Gamma_{Q}= 7.9\; Q^{5/2}$ \AA$^{-5/2}$. Increasing
the temperature well above the plateau in $\rm \kappa(T)$ has
a scarce efect on the frequencies of the longitudinal phonon
while its damping increases up to $\Gamma_{Q} = 13.9 \;
Q^{3/2}$ \AA$^{-3/2}$. The absence of a transverse
acoustic branch at such temperature is here ascribed to scattering from static
orientational disorder since, as shown below, such phonons are
observable in the high-temperature crystal where low-energy
librations are thermally populated. Increasing again the temperature up
to a value consistent with that explored experimentally, that is
0.88$\rm T_g$, leads to a substantial broadening of the
longitudinal mode frequencies for the ODC. These, as shown
in the lower frame of Fig. 2 correspond to a
hydrodynamic value for the sound velocity of 2349 m s$^{-1}$, a
result in agreement with light-scattering results \cite{eto1}. In
addition, the phonon remains as a well defined entity up to
wavevectors corresponding to one half of the Brillouin zone only,
and the damping coefficient can be accounted for as $\Gamma_{Q}= 62.40\; Q^{5/2}$.
Data for the monoclinic crystal also depicted in Fig.2 show a
longitudinal branch with a hydrodynamic limiting value of 3917 m
s$^{-1}$ together with a transverse mode propagating with a
hydrodynamic velocity of 1574 m s$^{-1}$. Damping of both modes is
well accounted for by hydrodynamic Q-dependences with coefficients
of 11.98 \AA$^2$ meV and 9.10 \AA$^2$ meV, respectively. Our data
contrast with the interpretation of IXS data on polycrystals given in Ref.\cite{matic}.
The technique,  as recently pointed out \cite{lb} mostly samples inhomogeneus spectra  
composed  by excitations additional to those of acoustic character \cite{ld} which
pre-empts the interpretation of maxima and linewidths of the spectral bands 
in terms of  physical frequencies and their damping terms.

\section{Discussion and Conclusions}

\begin{figure}
\includegraphics[width=3.0in,height=3.0in,angle=0]{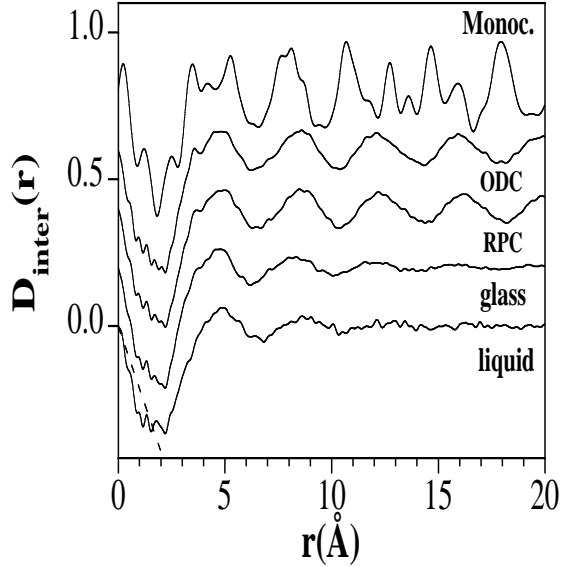} 
\caption{Experimental intermolecular static correlation functions 
$D(r) = 4 \pi \rho r [g(r)-1]$ derived from neutron diffraction 
measurements for all the condensed phases after subtraction of the 
molecular form-factor. Data for the monoclinic crystal and the ODC 
have been measured at 5 K. Data for the liquid correspond to T = 180 
K and those for the RPC to T = 105 K. The dotted line shows the 
density line $- 4 \pi \rho r$, where $\rho$ stands for the number 
density. Graphs are displaced 0.2 units upwards for display purposes. }
\label{Statics} 
\end{figure}

As a starting point of the discussion lets consider the signature of 
static correlations for all phases of ethanol as measured by 
means of neutron diffraction \cite{eto1}. The Fig.\ref{Statics} displays a set of 
intermolecular static correlation functions corresponding to all the 
condensed phases, including the normal liquid. A glance to the figure
shows, as expected, the close proximity between static correlations in 
liquid and glass phases as well as that existing between RPC and ODC
states. Interesting enough, one notices that the main differences 
between the functions corresponding to the disordered crystals and 
those for the liquid/glass states pertain the sustained oscillations
which appear having a period $p = 2\pi /Q_p$, where $Q_p$ stands for 
the position in momentum transfers where the main $bcc$ reflection is 
located. Notice however that the relative phases of such oscillations
in both fully disordered and $bcc$ samples are kept, irrespective of 
the presence of an underlying lattice.

In stark contrast, data pertaining the fully ordered crystal shows a 
far more complex pattern, which gets out-of-phase with those of the
disordered crystals for distances beyond those characteristic of the
first coordination sphere, that is above some 6 \AA - 7 \AA. For 
distances within such limit molecular orientational ordering arises
from strong electrostatic and packing effects, which are expected to
be not too different within the ODC and FOC states. One would then 
expect that excitations involving librational molecular motions would 
bear some resemblance in both crystals. 

\begin{figure}
\includegraphics[width=3.0in,height=3.0in,angle=0]{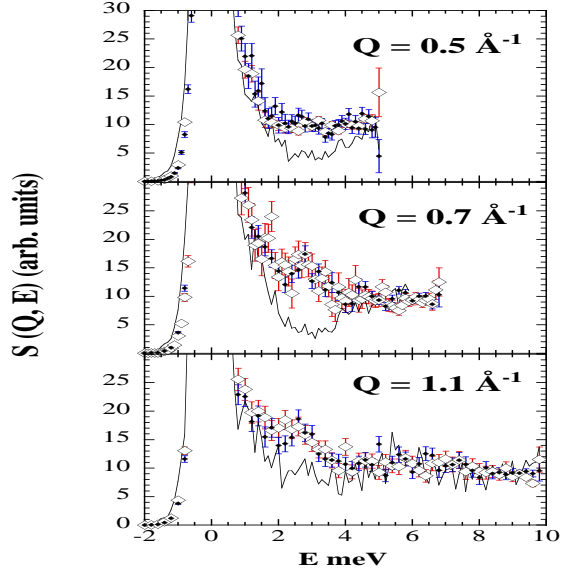} 
\caption{Inelastic neutron scattering spectrum for fully deuterated 
glassy (filled symbols) ODC (open symbols) and FOC (solid line) ethanol measured at T=10 K.}
\label{mari} 
\end{figure}

The lattice dynamics for the 
monoclinic crystal is well understood for frequencies up to about 25 
meV in terms of a semirigid molecular model \cite{ld}. The calculated 
frequency distributions compare favourably below such frequency 
threshold with those derived from neutron scattering for the fully
hydrogenated or deuterated samples \cite{ld}. The lattice dynamics
model gives rise to a set of 32 dispersion branches. Within those,
the acoustic modes are confined below some 8 meV, above which
lies a dense mesh of mildly dispersive modes. From the analysis of
the mode eigenvector components, it was found that purely acoustic 
modes dominate the frequency distribution below 4.2 meV and that 
the first well defined peak appearing at some 6 meV already has 
a significant optic character. The latter becomes dominant for
frequencies above 9 meV. In turn, rotational and librational 
motions within the $bcc$ crsytals have been considered in detail
(see Criado {\it et al.} in Ref.\cite{eto1}) and found to correspond
to reorientations amongst 24 different basic orientations set by
crystal site and molecular point symmetries. The frequency 
distribution of such motions covers a wide range extending up to 
about 20 meV. Furthermore, as shown by H.E. Fischer {\it et al.} in 
Ref.\cite{eto1}, the frequency spectra for both glass and ODC are
remarkably close and show below 5 meV the characteristic excess modes
of glassy matter. In comparison, the first van Hove singularity that
corresponds to zone boundary transverse acoustic phonons is still
seen at 80 K. Subsequent analysis of the spectral distributions
for both glass, ODC and FOC in terms of frequency moments 
$<\omega^n>$ has shown
how the lower order moments (up to $n=-1$), that are related to low frequencies 
and are thus pertinent for glass physics, for both glassy phases differ from that 
for the FOC, while higher order moments that weigth the higher 
frequencies become increasingly close for all the three phases. 
The facts just referred to serve to understand the close similitude
of results for glass, ODC and FOC reported on Ref.\cite{matic}. Such
similarity comes as a consequence of the closely related spectra of 
molecular orientational modes in all the three phases which dominates
the frequency range explorable by the IXS technique. However the
spectra for the glassy phases becomes strinkingly different from that
of the FOC within the freqenecy range dominated by the acoustic modes
as the Fig.\ref{mari} exemplifies. In other words, and to put things 
explicitly, the results reported on by in Ref.\cite{matic} are to be
taken as descriptive at best since the analysis there pursued 
where all the IXS intensities below some 22 meV are described in terms 
of a single physical frequency is not meaningful.    

\begin{figure}
\includegraphics[width=3.0in,height=3.0in,angle=0]{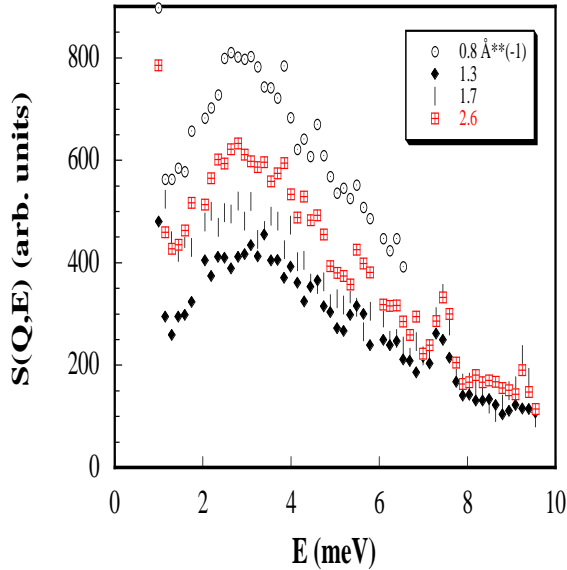} 
\caption{Inelastic neutron scattering spectrum for  
ODC ethanol measured at T=10 K as measured at several values of 
momentum-transfers and frequencies covering the maximum of the Boson 
peak. The peak at about 7.4 meV is a spurious instrumental reflection.}
\label{Boson} 
\end{figure}

Our final remark concerns the current debate on phonon localisation 
due to disorder \cite{silica}. In more detail, the discussion here 
concerns whether the propagation of acoustic excitations in glassy 
matter can take place beyond a spectral feature located at low 
frequencies known as the Boson peak. Its origin still is uncertain,
but is known to be related to the hump that appears in the specific
heat is plotted as $C_{p}/T^3$. Studies in vitreous silica \cite{lb} 
seems to indicate the existence of a crossover region well within 
the first pseudo-Brillouin zone that is for $Q<Q_p/2$ where the 
excitation linewidths equate their charateristic frequency and thus
cease to propagate. Such a cross-over frequency (or wavevector) 
usually referred to as the Ioffe-Regel limit has been identified in 
some studies with that correposnding to the Boson peak.  
For the materials under discussion 
here we show in Fig.\ref{Boson} the shape of the low-frequency spectra
for several values of momentum-transfers. Such a feature appears as a
well defined peak centered at about 3 meV which shows no clear Q-dependence
on its frequency or width. Its thermodynamic correlate is the 
$\approx$ T = 7 K hump in the specific heat reported in Ref.\cite{ld}.

Here we address the issue just referred to by means of a comparison of 
the simulation data depicted in Fig.\ref{FigMD} and the spectra
showing the Boson peak displayed in Fig.\ref{Boson}. Because of the
single-crystalline nature of the simulation data, the frequencies and
linewidths for the acoustic phonons cane be tracked down up to
the Brillouin zone boundary. Data for the 
ODC at the lowest temperature ( 5 K) shows no indication of a 
crossover in teh Ioffe-Reggel sense and therefore, one would expect
that longitudinal acoustic phonons propagate within the Brillouin zone. 
In contrast, data for T = 100 K do show a crossover for the ODC at 
about 0.4 \AA$^{-1}$. The inference that such a comparison allows us to 
make tells that thermal disorder seems a far more fundamental issue 
concerning the localization of acoustic waves than its static 
counterpart. The result, that went unnoticed in the current debate
is however well known concerning the damping of acoustic phonons 
in cubic crystals \cite{kuleev}, where it is kown that the phonon
relaxation rate depends on the fifth power of the temperature.

In summary, the concurrent use of thermal conductivity measurements, light scattering
experiments and molecular dynamics simulations on single crystalline 
models of the ODC and FOC solids has enabled us to establish the 
prominent role or molecular orientational disorder as the main
source of acoustic phonon scattering in disordered matter. As an
additional result, our study points towards the prominent role of 
thermal effects in the attenuation of acoustic phonons.

\end{document}